\documentclass[11pt]{article}

\usepackage[T1]{fontenc}
\usepackage[utf8]{inputenc}
\usepackage{lmodern}
\usepackage{microtype}
\usepackage[a4paper,margin=2.7cm]{geometry}
\usepackage[normalem]{ulem}
\usepackage{placeins}
\usepackage[numbers,sort&compress]{natbib}
\usepackage[colorlinks=true,allcolors=blue!55!black]{hyperref}
\usepackage{doi} 
\hypersetup{
  pdftitle={Clarifications on the Experimental Status of Real Quantum Theory},
  pdfauthor={Timothée Hoffreumon and Mischa P. Woods}
}

\usepackage{amsmath,amssymb,amsthm,mathtools}
\usepackage{bm}
\usepackage{xspace}
    \theoremstyle{plain}
    \newtheorem{theorem}{Theorem}

    \theoremstyle{definition}
    \newtheorem{definition}{Definition}

\usepackage{enumitem}
\setlist[itemize]{leftmargin=1.6em,itemsep=0.25em,topsep=0.4em}
\usepackage{nameref,cleveref} 
\crefname{equation}{Eq.}{Eqs.}
\Crefname{equation}{Eq.}{Eqs.}
\crefname{section}{Sec.}{Secs.}
\Crefname{section}{Sec.}{Secs.}
    \newlist{postrom}{enumerate}{2}
    \setlist[postrom,1]{label=\roman*.}
    \setlist[postrom,2]{label=\alph*. ref=\thepostromi\alph*}
    \crefname{postromi}{postulate}{postulates}
    \crefname{postromii}{postulate}{postulates}
    
    \newlist{postRom}{enumerate}{3}
    \setlist[postRom,1]{label=\Roman*. , ref=\Roman*}
    \setlist[postRom,2]{label=\alph*. , ref=\thepostRomi\alph*}
    \setlist[postrom,3]{label=\alph*. ref=\thepostRomii.\alph*}
    \crefname{postRomi}{Definition}{Definitions}
    \crefname{postRomii}{Definition}{Definitions}
    
    \crefname{post}{postulate}{postulates}

\usepackage{csquotes}
\usepackage{booktabs}
\usepackage{tikz}
\usetikzlibrary{arrows.meta,calc,positioning}
\usepackage{xcolor}

\newcommand{\nn}{\ensuremath{\mathbb{N}}}
\newcommand{\rr}{\ensuremath{\mathbb{R}}}
\newcommand{\cc}{\ensuremath{\mathbb{C}}}
\newcommand{\kk}{\ensuremath{\mathbb{K}}}

\newcommand{\tr}{\operatorname{tr}}
\newcommand{\trX}[2]{\tr_{#1}\left[ #2 \right]}
\newcommand{\id}{\mathbb{I}}

\newcommand{\KQT}{\ensuremath{\mathrm{(R)QT}}\xspace}
\newcommand{\KQTps}{\ensuremath{\mathrm{(R)QT}_{|\otimes}}\xspace}
\newcommand{\KQToi}{\ensuremath{\mathrm{(R)QT}_{|op.~ind.}}\xspace}
\newcommand{\QT}{\ensuremath{\mathrm{QT}}\xspace}
\newcommand{\QTps}{\ensuremath{\mathrm{QT}_{|\otimes}}\xspace}
\newcommand{\QToi}{\ensuremath{\mathrm{QT}_{|op.~ind.}}\xspace}
\newcommand{\RQT}{\ensuremath{\mathrm{RQT}}\xspace}
\newcommand{\RQTps}{\ensuremath{\mathrm{RQT}_{|\otimes}}\xspace}
\newcommand{\RQToi}{\ensuremath{\mathrm{RQT}_{|op.~ind.}}\xspace}

\newcommand{\mpwS}[1]{[MW:{\let\helpcmd\sout\parhelp#1\par\relax\relax}] }
\long\def\parhelp#1\par#2\relax{%
  \helpcmd{#1}\ifx\relax#2\else\par\parhelp#2\relax\fi%
}

\newcommand{\thS}[1]{{\let\helpcmd\sout\parhelp#1\par\relax\relax}}
\long\def\parhelp#1\par#2\relax{%
  \helpcmd{#1}\ifx\relax#2\else\par\parhelp#2\relax\fi%
}

\definecolor{reviewgreen}{RGB}{0,115,45}

\definecolor{penred}{RGB}{190,47,48}
\definecolor{penblue}{RGB}{39,94,157}
\definecolor{pengreen}{RGB}{46,111,64}
\definecolor{penmono}{RGB}{105,105,105}
\definecolor{bagfill}{RGB}{219,190,139}
\definecolor{panelblue}{RGB}{242,247,252}
\definecolor{panelgray}{RGB}{246,246,246}

\newcommand{\penicon}[5][]{%
  \begin{scope}[shift={(#2,#3)},rotate=#4]
    \draw[draw=black!82,line width=0.35pt,rounded corners=0.3pt,#1]
      (0,-#5) rectangle (1.03,#5);
    \draw[draw=black!82,line width=0.35pt,#1]
      (1.03,-#5) -- (1.34,0) -- (1.03,#5) -- cycle;
    \draw[draw=black!82,line width=0.35pt,#1]
      (-0.13,-#5) rectangle (0,#5);
  \end{scope}%
}

\newcommand{\personicon}[3][]{%
  \begin{scope}[shift={#2},scale=#3,line cap=round,line join=round,#1]
    \draw[fill=white] (0,1.75) circle (0.27);
    \draw[line width=0.8pt] (0,1.48) -- (0,0.62);
    \draw[line width=0.8pt] (0,1.23) -- (-0.48,0.88);
    \draw[line width=0.8pt] (0,1.23) -- (0.56,1.02);
    \draw[line width=0.8pt] (0,0.62) -- (-0.35,0);
    \draw[line width=0.8pt] (0,0.62) -- (0.38,0);
  \end{scope}%
}

\title{Clarifications on the Experimental Status of Real Quantum Theory}

\author{%
Timoth\'ee Hoffreumon\thanks{Electronic address: \href{mailto:t.hoffreumon@gmail.com}{t.hoffreumon@gmail.com}}\\
\small Mathematical Institute, Slovak Academy of Sciences, Bratislava, Slovakia
\and
Mischa P. Woods\thanks{Electronic address: \href{mailto:mischa.woods@gmail.com}{mischa.woods@gmail.com}}\\
\small ENS Lyon, Inria, Lyon, France
}

\date{\today}

\begin{document}

\maketitle

\begin{abstract}
In our previous paper on the matter, we proved that real quantum theory (RQT) cannot be experimentally falsified without the experiment falsifying quantum theory (QT) as well. 
Our result is in direct contradiction to the central claim of an article by Renou \emph{et al.}, published in \emph{Nature}~\textbf{600}, 625--629 (2021), reporting a bilocal network experiment that could falsify real quantum theory without falsifying quantum theory. 

The purpose of this note is to address several seemingly natural attempts to recover the experimental conclusion of Renou \emph{et al.} We first explain how these results, although based on nonlocality experiments, are far from a Bell-like level of proof. Then we explain why the proposed experiment requires an experimentally untestable premise about independent preparations necessarily being represented as product states. Finally, we discuss some concerns that lie outside the scope of both results. 
\end{abstract}

Renou \emph{et al.}~\cite{renou2021} present their proposal as an experimental falsification of real quantum theory (RQT), describing it as \emph{``...a Bell-like experiment, the successful realisation of which would disprove real quantum theory, in the same way as standard Bell experiments disproved local physics.''}

Our contribution~\cite{hoffreumon2026} identified two distinct assumptions underlying this conclusion. First, its claimed falsifying logic differs from Bell's: the bound on correlations is obtained only after restricting the states with which the systems enter the experiment: for the test to be conclusive, the sources must prepare product states in both the quantum theory (QT) and RQT descriptions. Second, the validity of this restriction cannot be experimentally certified within the framework considered. In particular, even if the product-state condition is accepted as a postulate, there is no procedure within the RQT framework that can certify that the prepared state satisfies it. Consequently, the proposed test, including its experimental implementations~\cite{Li_2022,Wu_2022}, rules out the stipulated product-source assumption but does not rule out RQT itself.

Section~\ref{sec:hypothetical_bell} presents our main argument and its contrast with Bell's theorem. Section~\ref{sec:untestable} examines the source definitions and certification problem; Sec.~\ref{sec:superdeterminism} addresses measurement independence and superdeterminism while Sec.~\ref{sec:fermionic} discusses superselected theories such as fermionic quantum theory. Appendices~\ref{sec:source_material} and~\ref{sec:certification} give the mathematical statements and the detailed certification argument.

\section{Why the proposed experiment does not falsify real quantum theory: an overview}\label{sec:hypothetical_bell}

Renou \emph{et al.}~\cite{renou2021} describe their proposal as an experimental falsification of real quantum theory analogous to the exclusion of local hidden-variable models by Bell experiments. To assess this analogy, we first recall what a Bell test excludes and then compare the preparation premises entering the two arguments.

In modern notation, a local hidden-variable model describes the observed conditional probabilities as
\begin{equation}\label{eq:LHV_intro}
p(a,b|x,y) = \int_{\Lambda}\mathrm{d}\lambda\,
\mu(\lambda)\,
p_A(a|x,\lambda)\,
p_B(b|y,\lambda),
\end{equation}
where $x,y$ are measurement settings, $a,b$ are outcomes, and $\mu$ is a normalized probability distribution over the hidden variable $\lambda$. The factorization of the response probabilities expresses Bell locality, while the setting independence of $\mu$ expresses measurement independence. Bell's theorem, and its CHSH formulation, constrain the entire class satisfying these assumptions.\footnote{Bell's original 1964 paper used a different correlation expression and inequality~\cite{bell1964}. For simplicity of exposition, we use the later CHSH formulation~\cite{clauser1969} throughout this comparison.} Writing $\mathcal{S}(\Lambda)$ for the CHSH value of the observed distribution $p(a,b|x,y)$ of the form \cref{eq:LHV_intro}, the bound $\mathcal{S}(\Lambda)\leq2$ holds for every allowed $\mu$ and every pair of local response functions. It does not require an additional restriction to a privileged class of source distributions~\cite{bell1964,clauser1969}. Quantum theory violates this bound and provides CHSH values of $\mathcal{S}(QT)\leq 2\sqrt{2}$.

The proposed experiment of Renou \emph{et al.} has a similar structure\footnote{See Appendix \ref{sec:source_material} for a comprehensive review of the argument.}, but with RQT taking the role of the candidate theory to be excluded, and the CHSH functional $\mathcal{S}$ replaced by the swapped-CHSH$_3$ functional $\mathcal{T}$. Measurements remain local to the laboratories: Alice measures $A$, Bob jointly measures the systems $B_1B_2$ in his laboratory, and Charlie measures $C$. Here \enquote{local measurements} does not mean that RQT satisfies Bell's hidden-variable factorisation; like QT, RQT can violate Bell inequalities. The distinction at issue concerns the additional source-preparation premise, not a relaxation of the locality of the measurements or of measurement independence. Relaxing measurement locality---for example, by allowing noncommuting measurements at different laboratories---only enlarges the class of admissible \RQT descriptions and cannot lower its maximal swapped-CHSH$_3$ value; it therefore cannot strengthen the claimed exclusion of \RQT.

In the Bell test, $\mathcal{S}(\Lambda)\leq 2$ holds for every source distribution allowed by the local hidden-variable model, whereas QT attains a maximal value of $2\sqrt{2}$. In the experiment of Renou \emph{et al.}~\cite{renou2021}, the corresponding real bound is $\mathcal{T}(RQT)\lesssim 7.66$, whereas QT attains a maximal value of $6\sqrt{2}\simeq 8.49$. Crucially, the real bound is obtained only after restricting the two-source input state to the product form
\begin{equation}
    \omega^{AB_1B_2C}=\rho^{AB_1}\otimes\sigma^{B_2C}.
    \label{eq:bell-comparison-product}
\end{equation}
There is no analogous restriction to a special subset of source distributions in the CHSH bound. Moreover, the product-source restriction is essential only on the RQT side of the proposed separation. As discussed in Sec.~\ref{sec:asymmetry}, the QT maximum is $6\sqrt{2}$ with or without this restriction: each conditional CHSH$_3$ expression is bounded by $6\sqrt{2}$, so their probability-weighted average obeys the same bound for arbitrary source states, while their product-source entanglement-swapping protocol attains it.

Calling the states in Eq.~\eqref{eq:bell-comparison-product} \enquote{independent sources} is a valid theoretical definition. Moreover, this representation may also be stipulated by postulate, as Renou \emph{et al.}~\cite{renou2021} do. However, accepting that stipulation does not establish that the actual laboratory preparation satisfies its antecedent; Sec.~\ref{sec:postulate} examines this distinction. The difficulty is establishing that the laboratory preparation belongs to this set. Within RQT, there are non-product states that give exactly the same statistics under all separate tests on the two sources as product states. We call states whose source-local statistics factorise \emph{operationally independent}. Product-state independence and operational independence thus define two subsets of the existing RQT state space, as explained in Sec.~\ref{sec:two-notions}; they are not competing postulates, and neither definition must be replaced by the other. Every product state is operationally independent, but the converse fails in RQT.

This distinction matters because the swapped-CHSH$_3$ separation disappears when the allowed input states are enlarged from product states to operationally independent real states. Our construction supplies non-product, operationally independent RQT preparations and local real measurements that reproduce the complete QT statistics of the protocol, including $\mathcal{T}[p]=6\sqrt{2}$~\cite{hoffreumon2026}. The claim is not that a fixed product state and a fixed non-product state remain indistinguishable under every possible joint measurement. Rather, source-local tests do not certify product structure, and the complete experiment admits both a QT description with product sources and an RQT description with operationally independent, non-product sources. This remains true when the protocol includes calibration, self-testing or global tomographic measurements: without an independently justified dimension bound, these procedures also admit an RQT explanation~\cite{hoffreumon2026}, as discussed in Sec .~\ref {sec:measurements} and formalised in Appendix~\ref{sec:certification}.

Observing the quantum value therefore does not have the unique explanation claimed by Renou~\emph{et al.}~\cite{renou2021}, namely that RQT has been experimentally excluded. An alternative explanation is that the preparation thought to consist of product-state-independent sources was instead operationally independent but non-product, with the experiment fully described by RQT. The violation excludes RQT together with the stipulated product-source premise; the data do not certify that premise. In a Bell test, by contrast, changing the source distribution within the local hidden-variable model cannot evade the CHSH bound. This contrast concerns the additional preparation restriction, not the absence of assumptions from Bell's theorem.


At first sight, one might think that this alternative is removed by stipulating that, in RQT, the joint description of independently prepared systems must be a product state by postulate. Unfortunately, this does not remove the problem because RQT also describes situations such as a faulty generation of independent sources, this is to say, source generators which actually produce non-product states. To ensure this is not the case, one could appeal to notions external to what the theory can describe, for example by identifying spacelike separation with a product-state description. However, because of the certification problem, such a premise has no experimentally accessible antecedent besides operational independence. As we mentioned above, this is at odds with the QT description, which does not have this problem: one could devise an experiment able to establish this product-state premise at least in principle. But more importantly, testing this premise has never been a necessity in QT since it does not affect the maximum value of $\mathcal{T}$ it predicts.

It is also worth noting that the RQT explanation only requires a mild form of bound entanglement. Therefore, it is not a form of  superdeterminism: the source state remains independent of subsequent measurement choices (Sec.~\ref{sec:superdeterminism}). Superselected quantum theories, including fermionic quantum theory, also admit experimentally indistinguishable RQT representations (Sec.~\ref{sec:fermionic}).

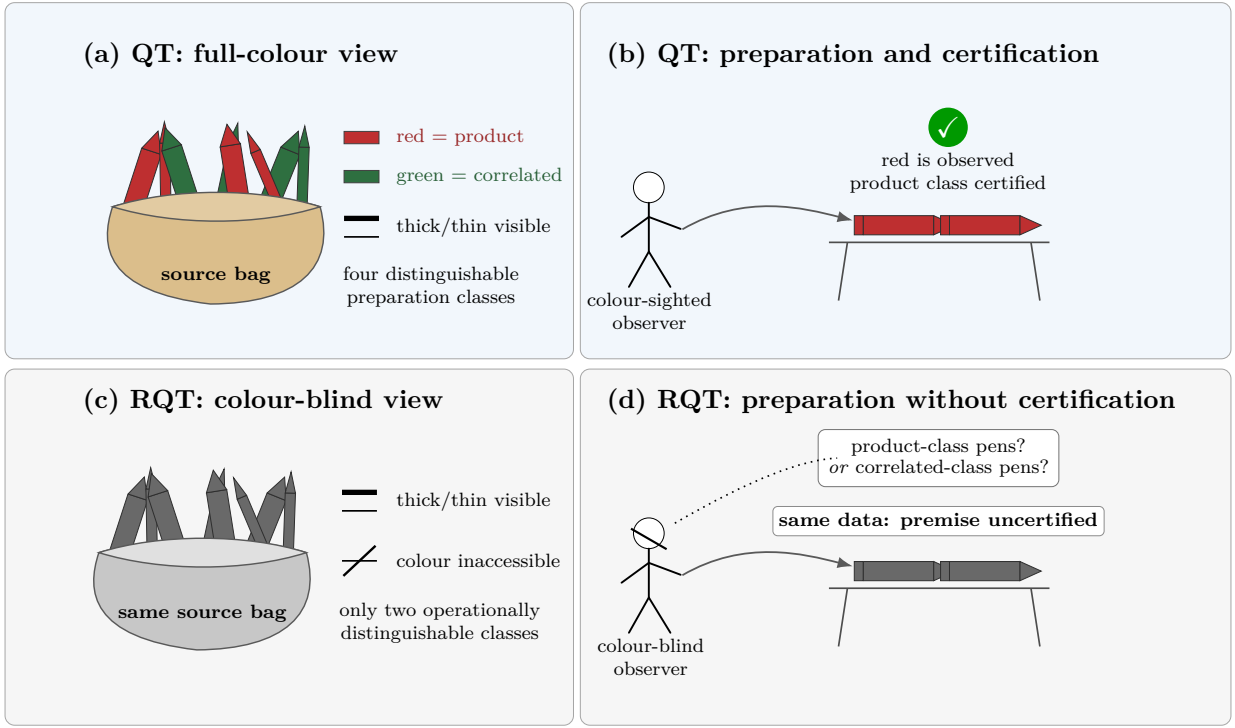
\begin{figure}[thb]
	\centering
	\makebox[\textwidth][c]{%
		\resizebox{1.04\textwidth}{!}{%
			\begin{tikzpicture}[font=\sffamily,>=Latex]
				\draw[rounded corners=4pt,fill=panelblue,draw=black!35]
				(-0.75,4.50) rectangle (7.55,9.70);
				\draw[rounded corners=4pt,fill=panelblue,draw=black!35]
				(7.65,4.50) rectangle (17.15,9.70);
				\draw[rounded corners=4pt,fill=panelgray,draw=black!35]
				(-0.75,-0.85) rectangle (7.55,4.35);
				\draw[rounded corners=4pt,fill=panelgray,draw=black!35]
				(7.65,-0.85) rectangle (17.15,4.35);
				
				\node[anchor=west,font=\bfseries] at (0.25,8.93)
				{(a) QT: full-colour view};
				\node[anchor=west,font=\bfseries] at (7.90,8.93)
				{(b) QT: preparation and certification};
				\node[anchor=west,font=\bfseries] at (0.25,3.88)
				{(c) RQT: colour-blind view};
				\node[anchor=west,font=\bfseries] at (7.90,3.88)
				{(d) RQT: preparation without certification};
				
				\begin{scope}[xshift=-3.5mm]
					\penicon[fill=penred]{1.42}{6.55}{72}{0.15}
					\penicon[fill=penred]{1.95}{6.62}{91}{0.075}
					\penicon[fill=pengreen]{2.35}{6.60}{108}{0.15}
					\penicon[fill=pengreen]{2.72}{6.63}{78}{0.075}
					\penicon[fill=penred]{3.05}{6.60}{99}{0.15}
					\penicon[fill=pengreen]{3.38}{6.58}{68}{0.15}
					\penicon[fill=penred]{3.66}{6.58}{113}{0.075}
					\penicon[fill=pengreen]{3.93}{6.56}{88}{0.075}
					\path[fill=bagfill,draw=black!65,line width=0.6pt]
					(1.17,6.72)
					.. controls (0.98,6.15) and (1.05,5.48) .. (2.60,5.30)
					.. controls (3.88,5.31) and (4.44,5.83) .. (4.20,6.72)
					.. controls (3.35,6.98) and (2.05,6.98) .. (1.17,6.72) -- cycle;
					\draw[fill=bagfill!80!white,draw=black!65,line width=0.6pt]
					(1.17,6.72) .. controls (2.08,6.44) and (3.32,6.44) .. (4.20,6.72)
					.. controls (3.33,7.00) and (2.05,7.00) .. (1.17,6.72) -- cycle;
					\node[font=\scriptsize\bfseries] at (2.68,5.72) {source bag};
					\draw[fill=penred,draw=black!75,line width=0.35pt]
					(4.55,7.64) rectangle (5.06,7.82);
					\node[anchor=west,font=\scriptsize,text=penred!80!black]
					at (5.18,7.73) {red = product};
					\draw[fill=pengreen,draw=black!75,line width=0.35pt]
					(4.55,7.07) rectangle (5.06,7.25);
					\node[anchor=west,font=\scriptsize,text=pengreen!85!black]
					at (5.18,7.16) {green = correlated};
					\draw[line width=2.2pt] (4.55,6.55) -- (5.06,6.55);
					\draw[line width=0.7pt] (4.55,6.28) -- (5.06,6.28);
					\node[anchor=west,font=\scriptsize] at (5.18,6.42)
					{thick/thin visible};
					\node[align=center,font=\scriptsize] at (5.82,5.55)
					{four distinguishable\\preparation classes};
				\end{scope}
				
				\personicon{(8.65,5.55)}{0.83}
				\node[font=\scriptsize,align=center] at (8.66,5.20)
				{colour-sighted\\observer};
				\draw[draw=black!70,line width=0.7pt] (11.28,6.20) -- (14.50,6.20);
				\draw[draw=black!70,line width=0.7pt] (11.55,6.20) -- (11.42,5.35);
				\draw[draw=black!70,line width=0.7pt] (14.23,6.20) -- (14.36,5.35);
				\penicon[fill=penred]{11.78}{6.45}{0}{0.14}
				\penicon[fill=penred]{13.04}{6.45}{0}{0.14}
				\draw[->,line width=0.75pt,draw=black!65]
				(9.14,6.39) .. controls (10.05,6.92) and (10.92,6.70) .. (11.63,6.52);
				\node[circle,fill=green!60!black,text=white,font=\bfseries\large,
				inner sep=1.2pt] at (13.02,7.88) {$\checkmark$};
				\node[align=center,font=\scriptsize] at (13.02,7.22)
				{red is observed\\product class certified};
				
				\begin{scope}[xshift=-5.5mm]
					\penicon[fill=penmono]{1.42}{1.50}{72}{0.15}
					\penicon[fill=penmono]{1.95}{1.57}{91}{0.075}
					\penicon[fill=penmono]{2.35}{1.55}{108}{0.15}
					\penicon[fill=penmono]{2.72}{1.58}{78}{0.075}
					\penicon[fill=penmono]{3.05}{1.55}{99}{0.15}
					\penicon[fill=penmono]{3.38}{1.53}{68}{0.15}
					\penicon[fill=penmono]{3.66}{1.53}{113}{0.075}
					\penicon[fill=penmono]{3.93}{1.51}{88}{0.075}
					\path[fill=white!78!black,draw=black!65,line width=0.6pt]
					(1.17,1.67)
					.. controls (0.98,1.10) and (1.05,0.43) .. (2.60,0.25)
					.. controls (3.88,0.26) and (4.44,0.78) .. (4.20,1.67)
					.. controls (3.35,1.93) and (2.05,1.93) .. (1.17,1.67) -- cycle;
					\draw[fill=white!88!black,draw=black!65,line width=0.6pt]
					(1.17,1.67) .. controls (2.08,1.39) and (3.32,1.39) .. (4.20,1.67)
					.. controls (3.33,1.95) and (2.05,1.95) .. (1.17,1.67) -- cycle;
					\node[font=\scriptsize\bfseries] at (2.68,0.79) {same source bag};
					\draw[line width=2.2pt] (4.72,2.55) -- (5.23,2.55);
					\draw[line width=0.7pt] (4.72,2.28) -- (5.23,2.28);
					\node[anchor=west,font=\scriptsize] at (5.38,2.42)
					{thick/thin visible};
					\draw[line width=0.7pt] (4.72,1.55) -- (5.23,1.55);
					\draw[line width=1.0pt] (4.74,1.33) -- (5.21,1.77);
					\node[anchor=west,font=\scriptsize] at (5.38,1.55)
					{colour inaccessible};
					\node[align=center,font=\scriptsize] at (6.15,0.65)
					{only two operationally\\distinguishable classes};
				\end{scope}
				
				\personicon{(8.65,0.50)}{0.83}
				\draw[line width=0.9pt] (8.39,2.02) -- (8.91,1.72);
				\node[font=\scriptsize,align=center] at (8.68,0.15)
				{colour-blind\\observer};
				\draw[draw=black!70,line width=0.7pt] (11.28,1.15) -- (14.50,1.15);
				\draw[draw=black!70,line width=0.7pt] (11.55,1.15) -- (11.42,0.30);
				\draw[draw=black!70,line width=0.7pt] (14.23,1.15) -- (14.36,0.30);
				\penicon[fill=penmono]{11.78}{1.40}{0}{0.14}
				\penicon[fill=penmono]{13.04}{1.40}{0}{0.14}
				\draw[->,line width=0.75pt,draw=black!65]
				(9.14,1.34) .. controls (10.05,1.87) and (10.92,1.65) .. (11.63,1.47);
				\node[draw=black!55,rounded corners=3pt,fill=white,align=center,
				font=\scriptsize,inner sep=4pt] at (12.88,3.05)
				{product-class pens?\\[-0.1em]\emph{or} correlated-class pens?};
				\draw[dotted,line width=0.7pt] (9.03,2.10) .. controls (10.15,2.75) and
				(10.90,3.03) .. (11.40,3.05);
				\node[draw=black!45,rounded corners=2pt,fill=white,
				align=center,font=\scriptsize\bfseries,inner sep=2.5pt] at (12.88,2.13)
				{same data: premise uncertified};
			\end{tikzpicture}%
		}%
	}
	\caption{Pen analogy for the distinction between preparation and certification.
		In both toy worlds [toy world 1: (a)--(b); toy world 2: (c)--(d)] the same analogue of Postulate~4 applies: {\it a preparation known
			to be independent is assigned to the red, product-state class.} Green represents non-product but operationally independent, i.e., correlated, preparations. Thickness represents an operational degree of freedom visible in both theories. In the full-colour \QT\ analogy, panels (a) and (b), the observer can distinguish all four combinations of colour and thickness. Two red pens can therefore be prepared and verified as red, allowing the postulate to identify the preparation with the product, uncorrelated class.  The presence of green pens emphasises that the postulate does not forbid correlated preparations in general, only that the preparation procedure was expected not to select them. 
		In the colour-blind \RQT\ analogy, panels (c) and (d), red pens remain valid objects, and two of them may certainly be prepared. However, red and green have the same accessible appearance: the colour-blind observer has no means to prevent themselves from selecting two green pens by mistake. Hence, while product preparation is not impossible, it cannot be established from the available observations.}
	\label{fig:pens}
\end{figure}

To finalize this section, a simple analogy captures the central issue at hand (Fig.~\ref{fig:pens}): Let red pens represent product-state preparations and green pens represent non-product, but operationally independent (hereafter `correlated') preparations. Pen thickness represents an ordinary degree of freedom that remains directly observable. A colour-sighted observer can therefore distinguish four kinds of pen---red or green, and thick or thin---and can both select and verify two red pens. A completely colour-blind observer can still distinguish thick from thin and may still physically select two red pens. The observer cannot, however, establish from the accessible appearance that the selected pens were red rather than green. The limitation concerns verification, not existence or physical possibility. In the analogy, the restricted visual data play the role of local RQT measurements, while the indistinguishability of the two colours within the available observational language corresponds to product states being locally indistinguishable from correlated states.

\section[Source independence and experimental certification]{Source independence and experimental certification}\label{sec:untestable}

\subsection{Two inequivalent definitions of source independence} We now examine the distinction between a source condition within the formalism and its certification in the laboratory, and explain why appeals to their Postulate~4 or additional preparation procedures do not close that gap.
\label{sec:two-notions}
Consider two sources whose joint state is denoted by $\omega_{ST}$. The independence condition imposed by Renou \emph{et al.}~\cite{renou2021} is
\begin{equation}
    \omega_{ST}=\rho_S\otimes\sigma_T,
    \label{eq:product-independence}
\end{equation}
or, when shared classical randomness is admitted, a convex mixture of such products.  We call \cref{eq:product-independence} \emph{product-state independence}.  It is a statement about the internal mathematical representative assigned to the joint preparation.

\begin{figure}[htbp]
\centering
\begin{tikzpicture}[font=\small]
  \node[font=\bfseries] at (-3.35,1.63) {Complex quantum theory};
  \draw[rounded corners=4pt,fill=black!2,draw=black!45]
    (-5.9,-1.23) rectangle (-0.8,1.28);
  \node[anchor=north west,font=\scriptsize,text=black!65]
    at (-5.72,1.13) {all joint states};
  \draw[fill=blue!10,draw=blue!60!black,line width=0.8pt]
    (-3.35,-0.15) ellipse (1.48 and 0.74);
  \node[font=\large] at (-3.35,-0.15)
    {$\mathsf P_{\mathbb C}=\mathsf O_{\mathbb C}$};

  \node[font=\bfseries] at (3.35,1.63) {Real quantum theory};
  \draw[rounded corners=4pt,fill=black!2,draw=black!45]
    (0.8,-1.23) rectangle (5.9,1.28);
  \node[anchor=north west,font=\scriptsize,text=black!65]
    at (0.98,1.13) {all joint states};
  \draw[fill=blue!10,draw=blue!60!black,line width=0.8pt]
    (3.35,-0.15) ellipse (1.78 and 0.82);
  \node[font=\large,text=blue!55!black]
    at (3.35,0.40) {$\mathsf O_{\mathbb R}$};
  \draw[fill=red!14,draw=red!65!black,line width=0.8pt]
    (3.05,-0.28) ellipse (0.80 and 0.35);
  \node[font=\large] at (3.05,-0.28) {$\mathsf P_{\mathbb R}$};
\end{tikzpicture}
\caption{Operational independence is an observational classification, not a
postulate.  The set $\mathsf O_{\mathbb K}$ consists of states locally
indistinguishable from the product of their marginals.  Local tomography makes
$\mathsf P_{\mathbb C}=\mathsf O_{\mathbb C}$ in \QT.  In \RQT,
$\mathsf P_{\mathbb R}\subsetneq\mathsf O_{\mathbb R}$; the additional real
states are non-product states with the same locally accessible statistics as
product states.}
\label{fig:state-sets}
\end{figure}
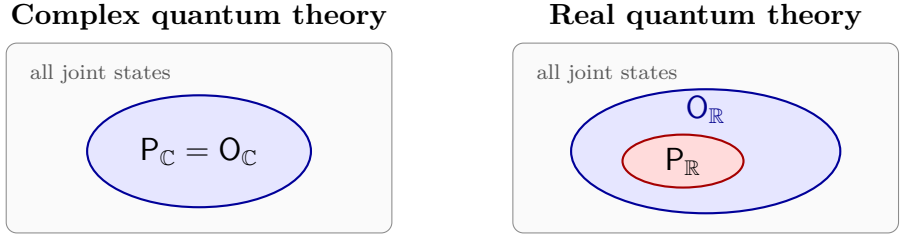

What a locality experiment can directly test is different.  Let $\{E_{a|x}\}_a$ and $\{F_{b|y}\}_b$ be arbitrary local measurements on the systems emitted by the two sources.  In addition to the above notion of independence, in Ref.~\cite{hoffreumon2026} we use another useful definition of a set of states, namely, sources are \emph{operationally independent} when
\begin{equation}
\begin{split}
 p(a,b|x,y)
 &=\tr\!\left[\omega_{ST}
        \bigl(E_{a|x}\otimes F_{b|y}\bigr)\right] \\
 &=p(a|x)\,p(b|y)
 \qquad\text{for all }a,b,x,y.
 \label{eq:operational-independence}
\end{split}
\end{equation}
Operational independence names a subset of the existing RQT state space. It is neither an additional postulate nor a replacement for their Postulate~4. It is only a name for an observationally defined subset of the state space: those joint states that no pair of allowed local measurements can distinguish from the product of their marginals.  If $\mathsf P_{\mathbb K}$ denotes the product states and $\mathsf O_{\mathbb K}$ the states satisfying \cref{eq:operational-independence} over the field $\mathbb K$, then $\mathsf P_{\mathbb K}\subseteq\mathsf O_{\mathbb K}$.  Local tomography gives equality for \QT, whereas the inclusion is strict for \RQT, as illustrated in Fig.~\ref{fig:state-sets}.

The failure is visible already for two rebits.  Define
\begin{equation}
 J=\begin{pmatrix}0&-1\\[0.15em]1&0\end{pmatrix},
 \qquad
 \Omega_r=\frac{1}{4}\bigl(\id\otimes\id+rJ\otimes J\bigr),
 \qquad |r|\leq 1.
 \label{eq:hidden-state}
\end{equation}
Since $J\otimes J$ is real and symmetric and has eigenvalues $\pm1$, $\Omega_r$ is a valid real density operator.  For $r\neq0$ it is not the product of its marginals, which are both maximally mixed.  Nevertheless, every local rebit effect is real symmetric, so $\tr(EJ)=0$ for every such effect $E$.  It follows that
\begin{equation}
 \tr\!\left[\Omega_r(E\otimes F)\right]
 =\frac{1}{4}\tr(E)\tr(F)
 =\tr\!\left[\frac{\id}{2}E\right]
  \tr\!\left[\frac{\id}{2}F\right].
 \label{eq:hidden-factorization}
\end{equation}
Thus $\Omega_r$ is non-product but operationally independent. The coefficient $r$ records a correlation that belongs to the real joint state but is invisible to all local real measurements. This is not a pathology added to the formalism: it is an elementary consequence of the failure of local tomography, a standard structural feature of real-vector-space quantum theory~\cite{wootters1990,caves2001,aleksandrova2013}.

The real representations constructed in Ref.~\cite{hoffreumon2026} use precisely this general possibility on arbitrary networks.  Our first theorem establishes that every finite network correlation obtainable in \QT\ has an \RQT\ realization with the same measurement-locality structure and with operationally independent sources.  The required real source state need not satisfy \cref{eq:product-independence}. Our second theorem extends the equivalence to arbitrary finite sequences of channels and measurements, each with a prescribed locality structure.  

The separation established by Renou \emph{et al.}~\cite{renou2021} therefore identifies a difference between \QT\ and the subclass of \RQT\ models in which the states of independent sources must be of the form of \cref{eq:product-independence}. However, it does not provide evidence as to why only this subclass of models can be used to describe the experiment. 

\subsection[What Postulate 4 establishes]{What Postulate 4 of Renou \emph{et al.} establishes}
\label{sec:postulate} For the purposes of this argument, we may accept that the assignment of product state to preparations classified as independent as a postulate. The question is whether the preparation implemented in the laboratory can be established to belong to that class. A postulate specifying its mathematical representation does not itself provide that experimental identification. 
Indeed, Renou \emph{et al.}~\cite{renou2021} present four postulates for the Hilbert-space formulation. Postulate~4 specifies tensor-product composition and the trivial action of operators acting on subsystems on the other subsystem, and then states: \emph{``Similarly, the state representing two independent preparations of the two systems is the tensor product of the two preparations.''}\footnote{Appendix~\ref{sec:KQT_def} isolates this clause as the product-source condition~\ref{PV_ps} to compare the relevant classes of models. Operational independence is introduced there as a definition of another subset of states, not as a replacement postulate.}
One might therefore object that independent sources are product states by definition, so the experiment needs no further certification.

This objection confuses a conditional statement inside a model with the identification of a laboratory event. In schematic form, the postulate asserts
\begin{equation}
 \text{preparations are independent}
 \quad\Longrightarrow\quad
 \text{joint representative is a product state}.
 \label{eq:conditional}
\end{equation}
This applies to both \QT and \RQT, without ever questioning if the identification is reasonable within their proposed experimental setup. In particular, it does not imply that spatial separation, separate controls, or the experimenters' intention to operate two devices independently guarantees that the actual joint real state has the form in \cref{eq:product-independence}; it postulates it. 

The logical point is especially transparent in the example of \cref{eq:hidden-state}. The product state $\id/2\otimes\id/2$ is a valid preparation, and so is the non-product state $\Omega_r$. Both give exactly the same statistics under every pair of local rebit measurements. Postulate~4 does not remove $\Omega_r$ from the real state space; rather, it imposes that the state $\Omega_r$ cannot be prepared by independent devices. This stipulation does not, however, establish that the experimental sources were product-state independent: they could instead have been operationally independent but non-product sources, indistinguishable by local measurements.

The remaining question concerns the application of the product-state requirement to the laboratory preparation, rather than its validity as a mathematical stipulation. It is useful to separate three propositions that are easily run together:
\begin{enumerate}[label=(\roman*),leftmargin=2.1em]
 \item the experimenters operate two (spacelike separated) devices using separate controls;
 \item all accessible cross-source tests yield factorised statistics;
 \item the theory assigns the joint preparation a product density operator.
\end{enumerate}
The first is a statement about laboratory organisation, the second is operational independence, and the third is product-state independence.  In \QT, background structural facts often allow one to move between (ii) and (iii) without comment, although see Sec.~\ref{sec:asymmetry} for when this is not the case.  In \RQT, that move is invalid. Treating it as automatic assumes precisely the property that the experiment is supposed to establish.

\subsection{Preparation is not certification}
Our argument in Ref.~\cite{hoffreumon2026} was not about operational independence being a `better definition' than product-state independence. Within the dimension-unbounded comparison considered here, the observed statistics do not certify that the intended product-source representation was implemented. The complete protocol also admits a non-product real realization. In other words, it showed the impossibility of finding a testable antecedent (ii) to bridge between the laboratory event (i) and its mathematical representation in the theory (iii).

Renou \emph{et al.}~\cite{renou2021} do state that the sources are independent and incorporate product form through Postulate~4. What they do not do is distinguish the laboratory arrangement (i) from the product-state requirement (iii), or show how experimental evidence establishes the latter without already assuming that QT, rather than RQT, describes the devices. This is not merely a hypothetical concern. Equations~\eqref{eq:hidden-state}--\eqref{eq:hidden-factorization} display valid real states for which the gap is exact, and the general Theorems~1 and~2 of Ref.~\cite{hoffreumon2026} show that it persists for the full protocols at issue.

The additional assumption is also particularly ill-suited to a test of \RQT. A fair comparison between theories must phrase the trusted physical conditions in terms that have the same operational meaning in each theory. Locality of separated measurements meets this requirement: it is captured by commutativity and by an imposed tensor-factor structure on the operations.\footnote{As in QT, these are indeed equivalent notions in RQT; see~\cite{fermionRQT}.} Product-state source independence does not. Because \RQT\ is not locally tomographic, it removes real states that are operationally indistinguishable from independent preparations.  Adopting it, therefore, selects a restricted foil theory before the experiment begins.

The simple analogy of Fig.~\ref{fig:pens} from Sec.~\ref{sec:hypothetical_bell} captures the distinction, where the mentioned product states are those of \cref{eq:hidden-factorization}.

\FloatBarrier
\subsection[Limits of experimental certification]{Limits of experimental certification}
\label{sec:measurements}

A second proposed resolution is to specify the preparation mechanism in greater detail. For example, one might say that a known projective measurement, an entanglement-breaking channel, or a trusted transformation prepares the two desired systems, and that the form of this operation guarantees a product output. This response merely transfers the disputed assumption from the state to the operation.

There are two possible levels of trust. If the exact QT description of the preparation device is assumed, including the features that exclude the RQT representation, then the conclusion is already contained in the device model.  One can write down the assumed matrices and determine theoretically whether they admit the desired restricted RQT representation; the laboratory statistics add no independent premise.
If, by contrast, the preparation device is characterised only through experimental tests, those tests become further channels and measurements in a larger protocol.  They are then covered by Theorem~2 of Ref.~\cite{hoffreumon2026}: the entire finite sequence has an \RQT\ description with the same observed statistics and the same locality structure at every stage.

One might instead appeal to global tomography.  Theorem~2 of Ref.~\cite{hoffreumon2026} already rules out this possibility: any finite tomographic scheme is itself a finite sequence of operations and measurements, and hence cannot certify product-state independence without contradicting the theorem.  The operational reason is also simple.  A finite informationally complete global measurement presupposes a known finite Hilbert-space dimension.  Renou \emph{et al.}~\cite{renou2021}, however, allow real realisations of arbitrary dimension, including arbitrarily large-dimensional ones. Recall that the untestability of the dimension~\cite{Brunner_2008} was the starting point of looking for falsification tests that do not assume a particular dimension or internal description of the devices. No fixed finite set of global measurements can therefore certify the reconstructed state against every higher-dimensional real realisation permitted by the claimed falsification result.  Invoking global tomography would add an unwarranted dimension bound rather than close the certification gap.

Global measurements can distinguish particular real product states from non-product states that yield identical statistics under all local real measurements, as illustrated in Appendix~D of Ref.~\cite{hoffreumon2026}. This does not, however, certify the experimental preparation without assuming the relevant state and measurement descriptions. The certification procedure must itself be included in the experimental protocol. Theorem~2 then ensures that, whenever the complete protocol admits a \QT\ description, it also admits an \RQT\ description with the same statistics. Adding such a procedure therefore cannot establish the claimed exclusion of \RQT.

In summary, we have found no observable condition to replace (ii) that would establish the product-state requirement (iii) without assuming a particular description of the devices. Adding certification procedures does not close this gap: as long as the complete protocol remains compatible with \QT\ and no dimension bound is imposed, the data cannot exclude a non-product \RQT\ description of the initial sources.
Sources in a product state can always be `spoofed' by sources in a non-product state having the same statistics for all local measurements. This is somewhat desirable, since a protocol capable of distinguishing between the product and spoofing states would provide a simple experiment falsifying QT (and all tomographically local theories).

\subsection{The asymmetric role of independent sources in QT and RQT}\label{sec:asymmetry}
One might object that, when the total Hilbert-space dimension is unknown, finite data need not distinguish a product source state from a correlated higher-dimensional state even in \QT. This observation is correct, but it does not restore the separation claimed by Renou \emph{et al.}~\cite{renou2021}.


Source correlations play asymmetric roles in the argument. Each conditional CHSH$_3$ expression is bounded by $6\sqrt{2}$, so their probability-weighted average obeys the same bound for arbitrary QT source states, as noted by Renou \emph{et al.}~\cite{renou2021}. A product-source implementation attains it. Allowing correlated sources therefore leaves the QT maximum unchanged.

Imposing a priori constraints on the sources, such as requiring them to be a product state, is therefore irrelevant to the \QT\ side of the comparison. By contrast, the real bound of Renou \emph{et al.}~\cite{renou2021} is obtained only after product-state independence is imposed on the \RQT\ sources; allowing valid correlated real source states removes the claimed separation. Thus, the failure to certify product structure leaves the maximal \QT\ value unchanged, but invalidates the restrictive premise required to infer a falsification of \RQT.

\section{Some final remarks}

\subsection{Allowing operationally independent states to represent independent preparations is not superdeterminism}
\label{sec:superdeterminism}

The presence of correlations between nominally distinct sources may invite an analogy with superdeterministic explanations of Bell experiments.  The analogy is mistaken.

In a measurement-dependent or superdeterministic account of Bell correlations, the hidden variable $\lambda$ is correlated with the later measurement settings, so that
\begin{equation}
 p(\lambda|x,z)\neq p(\lambda).
 \label{eq:measurement-dependence}
\end{equation}
The initial conditions must therefore encode, or otherwise be statistically coordinated with, the particular settings that the experimenters will choose. This abandons one of the assumptions behind Bell's theorem: measurement independence. In its strongest reading, it threatens the ordinary operational notion that the settings may be selected freely~\cite{bell1964,hall2010}.

No such dependence appears in our construction. The correlations required in the real initial state are fixed independently of the later measurement settings.  Theorem~2 of Ref.~\cite{hoffreumon2026} gives the stronger, context-independent statement: the map of the initial state depends only on the original initial state, and the map of each measurement or channel depends only on the corresponding operation.  Changing a future setting changes the mapped measurement associated with that setting; it does not require retuning the source state. The mapping from QT to RQT preserves the prescribed locality of operations and the independence of the initial state from subsequent measurement choices.

One may tell an optional cosmological story in which the hidden cross-source correlations descend from a common past, but nothing in the mathematical result requires such a story.  More importantly, even that interpretation would not correlate the source with the experimenters' future choices.  The construction requires correlations \emph{among source degrees of freedom}, not correlations \emph{between the source and the settings}.  It therefore neither removes experimental agency nor invokes the fine-tuning characteristic of a superdeterministic evasion of Bell's theorem.

\subsection{Superselected quantum theories}\label{sec:fermionic}
Does imposing a superselection rule, as in fermionic quantum theory, obstruct an experimentally indistinguishable real representation? Superselection rules restrict the allowed states and measurements within the complex Hilbert-space description. Our mapping applies to all states and measurements of unrestricted \QT, preserving their outcome statistics and the locality of the operations. It therefore also applies to the allowed subsets selected by such restrictions. Restricting the mapping to these subsets gives a corresponding real representation with the same experimental predictions. In particular, fermionic quantum theory admits an \RQT\ representation that preserves the locality of its physical operations and is experimentally indistinguishable from the original fermionic theory. Full details of the fermionic construction are given in Ref.~\cite{fermionRQT}.

\section{Conclusion}
The central distinction can be stated in one sentence: \emph{the ability to prepare a product state does not imply the ability to certify that a product state, rather than an operationally indistinguishable correlated real state, was prepared}.  Postulate~4 of Renou \emph{et al.} establishes the representation of an already identified independent preparation; it does not perform the identification.  Trusted or self-tested preparation devices do not help unless their decisive internal description is assumed, in which case the conclusion is built into the premise.  Hidden cross-source correlations do not entail superdeterminism because the real state is independent of future measurement choices.

These clarifications leave the mathematical bound of Renou \emph{et al.}~\cite{renou2021} intact within their stipulated model while removing the claimed experimental interpretation.  The lesson is broader than the particular case of real quantum theory.  When a foil theory has a different tomographic structure, familiar mathematical representations need not carry the same operational meaning.  An experiment can falsify a theory only through conditions that the experiment itself can, at least in principle, establish.

\section*{Acknowledgements}
We thank Jacopo Surace, Mirjam Weilenmann, Marc-Olivier Renou, Lucas Tendick, William K. Wootters, Titouan Carette, Stefano Pironio, Ognyan Oreshkov, Rob Spekkens, and Markus M{\"u}ller for helpful discussions that helped us identify which parts of our previous paper~\cite{hoffreumon2026} required further clarification in the present note.

AI utilisation statement: We used Grammarly to help with the redaction, and ChatGPT 5.6 Sol to help us identify and rewrite the parts of the text that were too terse.

\begingroup
\footnotesize
\setlength{\bibsep}{0.2em}
\bibliographystyle{unsrtnat}
\bibliography{operational_independence_rqt}
\endgroup

\newpage
\appendix
{\centering{\Huge \textbf{Appendix}}}
\section{Formal exposition of the source material}\label{sec:source_material}
This is a self-contained exposition of the material needed to understand the main text. It also serves as a mathematically precise formulation of the statements in the main text, should the reader ever need it.  
It is important to note that this section is chiefly concerned with the exposition of the mathematical results from sources~\cite{renou2021} and~\cite{hoffreumon2026}, not with whether or how these definitions and theorems are actually attached to a given experiment. That role is reserved for the main text and~\Cref{sec:certification}.

To stress this distinction, we use the wording \textit{definition} instead of \textit{postulate} to present the variations of the mathematical formalism used in both papers. We will promote a \textit{definition} to a \textit{postulate} in~\Cref{sec:certification} to indicate the additional step of asserting that such a mathematical definition applies to a concrete element of reality realised in an experiment. For example, when we define a Hilbert space $\mathcal{H}_S$ for a system $S$, we are merely stating that the label $S$ is called a system and that this label has some Hilbert space $\mathcal{H}_S$ associated with it. By contrast, when we postulate a Hilbert space $\mathcal{H}_S$ for a system $S$, we assert that the physical system identified as $S$ has a description that will involve the corresponding Hilbert space.

\subsection{Definitions of (Real) Quantum Theory}\label{sec:KQT_def} 

Both Renou~\emph{et al.}'s work~\cite{renou2021} and ours~\cite{hoffreumon2026} are in agreement with the mathematical definition of a local game. It is only when several sources are involved (what we have called a `multilocal game') that the choice of definitions needed to prove the seemingly contradicting results~\cite[Theorem 3]{renou2021} and~\cite[Theorem 1]{hoffreumon2026} differ. We shall now restate the minimal set of statements we agree on, as well as the two complementary definitions that separate our works. This minimal set is precisely what we mean when mentioning `(Real) Quantum Theory' (abbreviated \KQT) as a mathematical model in this article. With the `real' word added when the underlying Hilbert space is assumed real instead of complex.

\paragraph{Definitions needed to model (multi)local experiments.} When one says that a local scenario is described by (Real) Quantum Theory, they mean the following:
\begin{postRom}
    \item \label{PI} (State Space) 
    
    To every physical system $S$ with a finite number $d$ of degrees of freedom, there corresponds a Hilbert space $\mathcal{H}_S$ over $\kk$ of dimension $n\geq d$. 
    
    The set of states that the system $S$ can take corresponds to the set of positive, trace-one operators acting on this space. 
    
    \item \label{PII} (Measurement) 
    
    A (finite) measurement $\mathsf{E}^S$ of the system $S$ corresponds to any family $\{E_r^S\}_r$ of positive operators acting on $\mathcal{H}_S$ indexed by the value of a discrete random variable $r$ identified with a subset of $\nn$ (i.e., $r\in \{0, 1, ... \}$) and such that $\sum_r E_r^S = \id^S$.
    
    \item \label{PIII} (Born rule) 
    
    If we measure $\mathsf{E}^S$ when the system $S$ is in state $\rho_S$, the probability of obtaining result $r$ is given by the inner product of the corresponding operators, $p(r) = \trX{}{\rho_S \: E_r^S }$. 
    
    \item \label{PIV} (Subsystems are tensor factors) 

    The Hilbert space $\mathcal{H}_{ST}$ corresponding to the composition of two subsystems $S$ and $T$ is the space $\mathcal{H}_S \otimes \mathcal{H}_T$.

    The action of only measuring subsystem $S$, i.e., performing a \textit{local measurement} of $S$, is represented by a measurement $\mathsf{E}^{ST}$ necessarily of the form $\{E_r^S \otimes \id^T\}_r$.

    The composition of performing a local measurement $\mathsf{E}^{ST}= \{E_r^S \otimes \id^T\}_r$ of $S$ with a local measurement $\mathsf{F}^{ST}= \{\id^S \otimes F_q^T\}_q$ of $T$ is given by the measurement $(\mathsf{EF})^{ST} = \{E_r^S \otimes F_q^T\}_{r,q}$ indexed by the value of the pair $(r,q)$ identified with a subset of $\nn \times \nn$.
    
\end{postRom}
In the above, $\kk$ denotes either the real or complex numbers, i.e., whether these statements define Quantum Theory or Real Quantum Theory; $\id^S$ (respectively, $\id^T$) denotes the identity operator acting on $\mathcal{H}_S$ (resp., $\mathcal{H}_T$). 

Keep in mind that these definitions are far from sufficient to encapsulate all of the quantum formalism. Also, note that all dimensions and cardinalities are assumed finite to simplify the discussion. Nevertheless, this is all that is needed to model several spacelike-separated parties locally measuring parts of a shared quantum system with a finite number of degrees of freedom---picture it as experimentalists at various places on Earth simultaneously observing the passage through their labs of one of the decay products of a particle that disintegrated at an earlier time and each of the experimentalists measures the spin of the decay product they saw. This is the situation we refer to as a `local experiment' and the setup for traditional Bell nonlocality.

When several sources distribute systems among the parties, the resulting network scenario can be analysed under different restrictions on the initial source state. Spatial separation of the source devices is a feature of the laboratory arrangement; whether it justifies a product-state representative is the certification question discussed in the main text. 
For our purposes of comparing two theorems, we define two source conditions within the quantum formalism. The labels \ref{PV_ps} and \ref{PV_oi} below are retained for ease of reference; they identify restrictions on the source states, since in either case, the states in the alternative class are still valid states within the theory, they are just not ones defined as `independent'. 
\begin{postRom}[resume]
    \item \label{PV}
    \begin{postRom}
        \begin{subequations}
        \item \label{PV_ps} (Independently prepared subsystems are product states)

       If \mbox{$S$} and \mbox{$T$} are independently prepared subsystems, the joint system is represented by a state \mbox{$\omega_{ST}$} necessarily of the form
        \begin{equation}\label{eq:PV_ps}
            \omega_{ST} = \rho_S \otimes \sigma_T \:, 
        \end{equation}
        where $\rho_S$ and $\sigma_T$ are valid states of their respective subsystems $S$ and $T$.

        \item \label{PV_oi} (Independently prepared subsystems lead to independent distributions)

        If \mbox{$S$} and \mbox{$T$} are independently prepared subsystems, the joint system is represented by a state \mbox{$\omega_{ST}$} that must yield independent statistics for all local measurements of \mbox{$S$} and \mbox{$T$}.
        
        That is, for any pair of local measurements $ \mathsf{E}^{ST}=\{E_r^S \otimes \id^T\}_r$ and $\mathsf{F}^{ST}=\{\id^S \otimes F^T_q\}_q$ such that $p(r,q) = \trX{}{ \omega_{ST} (E_r^S \otimes F_q^T)}$, the following should hold:
        \begin{equation}\label{eq:PV_oi}
            \exists~ p(r),\: \exists~ p(q): \qquad p(r,q) = p(r)p(q) \quad \forall r, \: \forall q\:.
        \end{equation}
        \end{subequations}
    \end{postRom}
\end{postRom}

\Cref{PV_ps} was used to derive the mathematical results of Ref.~\cite{renou2021} leading to their Theorem 3 restated below, whereas \cref{PV_oi} was used to derive the Theorem 1 of Ref.~\cite{hoffreumon2026}, also restated below. 
The results of both works will be presented assuming the corresponding definitions in the next sections. As such, the opposite conclusions reached by these two papers appear to follow from a mere definition issue. The goal of the rest of this paper is to argue that it is not. 

\begin{definition}
    When seen only as a mathematical definition, 
    \begin{itemize}
        \item \textbf{Quantum-like Theories} (abbreviated by `\KQT'), refers to theories obeying \Cref{PI,PII,PIII,PIV} for $\kk$ unspecified;
        \item \textbf{Quantum Theory} (`\QT'), refers to the theory obeying \Cref{PI,PII,PIII,PIV} for $\kk = \cc$;
        \item \textbf{Real Quantum Theory} (`\RQT'), refers to the theory obeying \Cref{PI,PII,PIII,PIV} for $\kk = \rr$; 
        \bigskip
        \item \textbf{Quantum Theory with product-state independent sources} (`\QTps'), refers to the theory obeying \Cref{PI,PII,PIII,PIV,PV_ps} for $\kk = \cc$;
        \item \textbf{Quantum Theory with operationally independent sources} (`\QToi'), refers to the theory obeying \Cref{PI,PII,PIII,PIV,PV_oi} for $\kk = \cc$;
        \bigskip 
        \item \textbf{Real Quantum Theory with product-state independent sources} (`\RQTps'), refers to the theory obeying \Cref{PI,PII,PIII,PIV,PV_ps} for $\kk = \rr$;
        \item \textbf{Real Quantum Theory with operationally independent sources} (`\RQToi'), refers to the theory obeying \Cref{PI,PII,PIII,PIV,PV_oi} for $\kk = \rr$.
    \end{itemize}
\end{definition}
For the quantum case, i.e. for $\kk=\cc$, \cref{PV_ps,PV_oi} are equivalent, whereas for $\kk=\rr$, there are more operationally independent states than product states~\cite{hoffreumon2026}. 

Because there are fewer states qualifying as independent in \RQTps than in \RQToi, for some multilocal scenarios, the set of all possible distributions predicted when assuming \cref{PI,PII,PIII,PIV,PV_ps} is provably smaller than the set obtained under \cref{PI,PII,PIII,PIV,PV_oi} for $\kk = \rr$ (an example will be provided below). This fact will be expressed concisely as 
\begin{equation}\label{eq:inclusion_RQTps_RQToi}
    \RQTps \subset \RQToi \:.
\end{equation}
For $\kk=\cc$, the two source conditions define the same set of states, so the corresponding classes of distributions coincide. Accordingly, we write:
\begin{equation}\label{eq:inclusion_QTps_QToi}
    \QTps = \QToi \:.
\end{equation}

The interest in studying theories phrased on an arbitrary number field $\kk$ is to check if every quantum prediction has a real quantum counterpart~\cite{Gisin_2007}. Actually, it has been known for a while that every outcome distribution obtainable in a local experiment described by quantum theory can equivalently be described by real quantum theory, provided the dimensions of the Hilbert spaces are suitably extended~\cite{Pal_2008,Moroder_2013}. 
We write this concisely as:
\begin{equation}
    \QT \subseteq \RQT \:.
\end{equation}

With this kind of notation, the main technical point of Renou \emph{et al.} can be phrased as follows~\cite{renou2021}:
\begin{equation}
    \QTps \supset \RQTps \:.
\end{equation}
That is, for some multilocal scenario, there are outcome distributions predicted by QT with product-state independence that cannot be predicted by RQT with product-state independence.
The dependence on a source restriction is explicitly acknowledged in Ref.~\cite{Weilenmann2025}, which quantifies the real source entanglement needed to reproduce the observed correlations and makes its exclusion of \RQT\ conditional on sufficiently limiting that entanglement. What remains unestablished is how the experimental preparation can be certified to satisfy this restriction. Quantifying the correlations required by an \RQT\ explanation does not exclude that explanation: the same observations remain compatible with product-source \QT\ and non-product \RQT\ descriptions, as discussed in \Cref{sec:measurements,sec:certification}.

Related real constructions were considered in Refs.~\cite{Barrios2025,Hoffreumon2025}. Ref.~\cite{hoffreumon2026} subsequently established that the class of distributions obtainable with operationally independent real sources contains its complex counterpart, with strict inclusion. 
\begin{equation}
    \QToi \subset \RQToi \:.
\end{equation}

Putting everything together, we can say that, at a purely technical level, the following chain of inclusions is true:
\begin{equation}
    \RQTps \subset \QTps = \QToi \subset \RQToi \:.
\end{equation}
The rest of this section is a presentation of the actual theorems underlying these inclusions, adapted to the particular network structure proposed by Renou \emph{et al.} .  

\subsection[Real representation of local experiments]{Real representation of local experiments}
We begin with some notation, nomenclature, and considerations which are customary in the nonlocality literature and which equally apply to \KQT. As is customary in quantum nonlocality, the variable $r$ indexing the measurement $\mathsf{E}^S$ is seen as a discrete random variable called the \textit{outcome} of this measurement, and the operators $E_r^S$ indexed by $r$ are called the \textit{effects}. 
The measuring party can consider more than one measurement at once, provided their respective outcome spaces are properly identified. When this is the case, the set $\{ \mathsf{E}^S_{|q}\}_q$ of possible measurements considered by the party is indexed by another discrete random variable $q$ called the \textit{setting}, and also identified with a subset of $\nn$ like $\{ \mathsf{E}^S_{|q}\}_q = \{ \{ E_{r|q=0}\}_r, \{E_{r|q=1}\}_r, ... \}$. 
The value of the random variable $r$ is thus conditioned on the value of another random variable $q$, and the Born rule gives rise to conditional densities like, e.g., $p(r|q) = \trX{}{E^S_{r|q} \: \rho_S}$ which is to be read as `the probability of observing outcome $r$ when the system $S$ in a state $\rho$ was measured by $\mathsf{E}$ in the setting $q$'.

When several local measurements on different subsystems are performed, the probability distribution obtained via the Born rule is multivariate, with each random variable associated with the outcome of its corresponding local measurement. These random variables are also local in the sense that they are accessible only to the party performing the conditional measurement that produced them. Similarly, when local measurements are conditioned by measurement settings, each setting is assumed to be local in the sense of accessible only to the party performing the conditional measurement associated with that setting. However, local settings are furthermore assumed to be freely chosen: the setting of one local measurement is never correlated to that of another local measurement (unless otherwise specified).

For example, for a composite system $AB$, let there be one party locally measuring subsystem $A$ via the measurement $\{ \mathsf{A}_{|x}\}_x$ depending on a setting $x$ and resulting in an outcome $a$\footnote{As is customary in quantum nonlocality, when there is no risk of confusion, the sans-serif letter associated with a measurement, the capital font letter associated with a measurement effect, and the small font letter associated with the outcome will all be taken as the same as the capital font letter associated with the subsystem---in this case since it is subsystem $A$, the measurement $\{ \mathsf{A}_{|x}\}_x$ is a shorthand for $\{ \mathsf{A}_{|x}^A = \{A^A_{a=0|x}, A^A_{a=1|x}, ...\}_a\}_x$.}, and let there be another party locally measuring subsystem $B$ via the measurement $\{ \mathsf{B}_{|y}\}_y$ depending on setting $y$ and resulting in an outcome $b$. According to \cref{PIV}, the joint measurement performed by these two parties is given by $\{ \mathsf{A}_{|x} \otimes \mathsf{B}_{|y} \}_{x,y}$; here it is indexed by the pair of settings $(x,y)$, and so the measurement effects are operators of the form $ A_{a|x} \otimes B_{b|y}$ acting on $\mathcal{H}_A \otimes \mathcal{H}_B$. According to \cref{PIII}, the joint probability distribution of outcomes $a$ and $b$ conditioned by the settings $x$ and $y$ is given by
\begin{equation}
    p(a,b|x,y) = \trX{}{\omega_{AB} (A_{a|x} \otimes B_{b|y})} \:,
\end{equation}
when the joint system $AB$ was in a state $\omega_{AB}$. 

In the above equation, notice how the local measurement on $B$ has no dependency on $x$ nor $a$, and, similarly, notice how the local measurement on $A$ has no dependency on $y$ nor $b$. This is what we meant by the outcomes and settings being accessible only to the local(ly measuring) party. 

For the purposes of this exposition, the discussion will be restricted to a local experiment involving four subsystems shared among three measuring parties, hereafter called `Alice', `Bob', and `Charlie'. Alice, Bob, and Charlie can thus be pictured as each sitting in a local, spacelike separated lab. Alice and Charlie each receive a subsystem, respectively denoted by $A$ and $C$, whereas Bob receives two, respectively denoted by $B_1$ and $B_2$. In addition, Alice and Charlie respectively receive the independently sampled settings $x$ and $z$ upon which their measurement may depend. 

Assuming that the parties each measure their subsystem(s) in order to produce the outcomes $a$, $b$, $c$ (Bob is performing a joint measurement, treating the Hilbert space $\mathcal{H}_{B_1} \otimes \mathcal{H}_{B_2}$ as a single space $\mathcal{H}_{B}$), and assuming that the experiment is described by \KQT, the distribution of outcomes given settings has the form
\begin{equation}\label{eq:bilocal_local}
    p(a,b,c|x,z) = \trX{}{\omega_{AB_1B_2C}(A_{a|x} \otimes B_b \otimes C_{c|z})} \:.
\end{equation}
Here, the sample spaces of the settings and outcomes, as well as the dimension of each subsystem, are yet to be specified.

Suppose the parties trust the outcome distribution, but not the theory that yielded the distribution. Is there a way for them to differentiate whether the distribution was obtained in accordance with \QT or \RQT? 
The immediate answer would be yes: since the (sub)system(s) may require more Hilbert-space dimensions to be represented over the reals than over the complexes, the parties could devise a protocol to check the dimension and conclude according to which theory the systems behaved.
However, such protocols are known not to exist~\cite{Brunner_2008}. This actually makes sense: if the parties do not trust the theory, why should they trust a quantity that only made sense within the theory?

As we mentioned above, any distribution $p(a,b,c|x,z)$ resulting from a local experiment modelled by \QT can also be modelled by \RQT.
\begin{theorem}[Adapted from Refs.~\cite{Pal_2008,Moroder_2013}] \label{thm:QT_subseteq_RQT}
    For a distribution of the form \cref{eq:bilocal_local} obtained via a set of complex operators $\{\omega_{AB_1B_2C},\: A_{a|x},\: B_b,\:  C_{c|z}\}$, there always exists a set of real operators $\{\Tilde{\omega}_{AB_1B_2C},\: \Tilde{A}_{a|x},\: \Tilde{B}_b,\:  \Tilde{C}_{z|y}\}$ such that 
    \begin{equation}
        \trX{}{\omega_{AB_1B_2C}(A_{a|x} \otimes B_b \otimes C_{c|z})} = \trX{}{\Tilde{\omega}_{AB_1B_2C}(\Tilde{A}_{a|x} \otimes \Tilde{B}_b \otimes \Tilde{C}_{z|y})} \:.
    \end{equation}
\end{theorem}
Therefore, the parties have virtually no way to tell \QT apart from \RQT for this kind of local experiment, at least as long as no violation of the predictions of \QT is observed. 

\subsection[Real representation of bilocal experiments]{Real representation of bilocal experiments}
But to find a separation in the predictions of \QT and \RQT, one might consider a variation of the game where the set of states the sources can be in is restricted. This is where \cref{PV} must be leveraged: the four systems are not assumed to come from a common source, but from two independent sources: one outputting systems $A$ and $B_1$ and the other outputting systems $B_2$ and $C$. 
Bob's measurement of the joint system $B_1B_2$ is thus the only thing able to induce correlations between Alice's and Charlie's measurements. This scenario is called a \textit{bilocal} experiment (of network nonlocality)~\cite{Branciard2010}.

In this restricted class of scenarios, the conditions for applying \Cref{thm:QT_subseteq_RQT} are no longer verified, and the question of whether all predictions of QT can be explained with RQT reopens. As we have seen, there are two possible interpretations of how to restrict the set of all states of the sources to represent their independence, \cref{PV_ps,PV_oi}.

When the underlying prepared state is assumed to be in \KQTps, the outcome distribution of a bilocal experiment is modelled by
\begin{equation}\label{eq:bilocal_ps}
    p_{|\otimes}(a,b,c|x,z) = \trX{}{(\rho_{AB_1} \otimes \sigma_{B_2C})(A_{a|x} \otimes B_{b} \otimes C_{c|z})}\:,
\end{equation}
where $\rho_{AB_1}$ is the state of the first bipartite source and $\sigma_{B_2C}$ is the one of the second source.
When the underlying prepared state is assumed to be in \KQToi instead, the outcome distribution is modelled by 
\begin{subequations}\label{eq:bilocal_oi}
\begin{equation}
    p_{|\otimes}(a,b,c|x,z) = \trX{}{\omega_{AB_1B_2C}(A_{a|x} \otimes B_{b} \otimes C_{c|z})}\:,
\end{equation}
where the joint state of the two sources must verify the condition
\begin{equation}
\begin{gathered}
    \begin{aligned}
    &\forall \mathsf{E}^{AB_1} = \{E^{AB_1}_r\}_r, \: \forall \mathsf{F}^{B_2C} = \{F^{B_2C}_q\}_q, \text{ s. th. } p(r,q) = \trX{}{\omega_{AB_1B_2C} (E^{AB_1}_r \otimes F^{B_2C}_q)}; \\
    & \exists p(r), \: \exists p(q), \text{ s. th. }
    \end{aligned}\\
    p(r,q) = p(r)p(q) \:.
    \end{gathered}
\end{equation}
\end{subequations}
Each choice will lead to a different conclusion. 

\subsubsection{Our first theorem}
We start by presenting an adaptation of the first result of our work~\cite{hoffreumon2026} to the bilocal game. It proves that all distributions predicted by \QToi can also be predicted by \RQToi. This will be relevant to understand the result of Renou~\emph{et al.}~\cite{renou2021}. 
\begin{theorem}[Adaptation of Theorem 1 from Ref.~\cite{hoffreumon2026}] \label{thm:QToi_subseteq_RQToi}
    For a distribution of the form \cref{eq:bilocal_oi} obtained via a set of complex operators $\{\omega_{AB_1B_2C},\: A_{a|x},\: B_b,\:  C_{c|z}\}$ acting on a Hilbert space of (complex) dimension $d_Ad_{B_1}d_{B_2}d_C$, there always exists a set of real operators $\{\Tilde{\omega}_{AB_1B_2C},\: \Tilde{A}_{a|x},\: \Tilde{B}_b,\:  \Tilde{C}_{z|y}\}$ acting on a Hilbert space of (real) dimension $6d_Ad_{B_1}d_{B_2}d_C$ such that 
    \begin{equation}
        \trX{}{\omega_{AB_1B_2C}(A_{a|x} \otimes B_b \otimes C_{c|z})} = \trX{}{\Tilde{\omega}_{AB_1B_2C}(\Tilde{A}_{a|x} \otimes \Tilde{B}_b \otimes \Tilde{C}_{z|y})} \:.
    \end{equation}
\end{theorem}
This result generalises to any multilocal scenario, that is, to any number of independent sources and local measurements and to any topology of the network (i.e., which subsystem is measured by whom); see the source material for the full proof. Hence, the inclusion $\QToi \subseteq \RQToi$.

\subsubsection{The proposed experiment of Renou~\emph{et al.}}

Rather than directly tackling the problem for \RQTps, Renou \emph{et al.} instead considered a Bell-like argument. To show \QTps and \RQTps lead to different predictions, they studied a functional from the set of achievable distributions to the positive reals. The separation is proven by showing the maximal value attainable with \QTps to be strictly larger than the maximal value attainable with \RQTps.

In more detail, they adapted the `CHSH$_3$' local game~\cite{Acin2016,Bowles2018} into a bilocal game. In this adaptation, Alice receives a setting $x\in\{0,1,2\}$ and produces an outcome $a\in\{0,1\}$, Bob produces an outcome $b\in \{0,1,2,3\}$, and Charlie receives a setting $z\in\{0,...,5\}$ and produces an outcome $c\in \{0,1\}$. The goal of the players is to obtain the highest value of a Bell-like functional $\mathcal{T} = \sum_{a,b,c} \:t(a,b,c)\: p(a,b,c|x,z)$ (the exact form of the function $t(a,b,c): \{0,1\}\times\{0,1,2,3\}\times \{0,1\} \rightarrow \{1,-1\}$ can be found in Ref.~\cite{renou2021}, but it is not of relevance for the discussion). 
To do so, they can decide prior to the experiment which measurements they will apply depending on the received setting, and which states the two shared systems must be prepared in. 

We call this game the `swapped CHSH$_3$ game' since the optimal quantum strategy to win the game is for Alice and Charlie to play the optimal strategy for the CHSH$_3$ game on their systems while Bob performs entanglement swapping on his~\cite{renou2021}. We thus refer to $\mathcal{T}$ as the `swapped CHSH$_3$ bound'. 

The interest in considering such a Bell-like functional lies in how its maximal value is not only dependent on the strategy adopted by the parties, but also on the description of sources and measurements. The functional can be defined regardless of the theory assumed to describe how the probability distribution was obtained, but each theory may predict a different maximal value it can take. In that regard, it provides a theory-independent criterion to test a theory: instead of asking ``can the predictions of this theory be reproduced with that other theory?'', it reduces the question to ``can the maximal value predicted by this theory be reached by that other theory?''.

In the following, we shall write $\mathcal{T}(\mathrm{<theory>})$ to concisely mean `the values the swapped-CHSH$_3$ functional can take for sources and measurements described by $\mathrm{<theory>}$'.

\begin{theorem}[Adapted from Ref.~\cite{renou2021}]\label{thm:Renou}
    Let $\mathcal{T}$ be the swapped-CHSH$_3$ functional obtained from a distribution of the form \cref{eq:bilocal_ps}.  Then
    \begin{enumerate}
    	\item Assuming \QTps, i.e. \cref{PI,PII,PIII,PIV,PV_ps} for $\kk=\cc$, and assuming every subsystem is associated with a 2-dimensional Hilbert space, the value of $\mathcal{T}$ is upper bounded by $6\sqrt{2}$.
    	
    	\item Assuming \RQTps, i.e. \cref{PI,PII,PIII,PIV,PV_ps} for $\kk=\rr$, the value of $\mathcal{T}$ is upper bounded by $7.66$, regardless of the dimension of the Hilbert spaces associated with each subsystem\footnote{Actually, their analysis goes a bit beyond \cref{PV_ps} as they also considered separable states to represent independently prepared sources. Nevertheless, this extension of the set of allowed states had no impact on the bound~\cite{renou2021,Weilenmann2025}, so there is no need to discuss it here.}.
    \end{enumerate}
\end{theorem}
We can re-express this theorem concisely as the bounds
\begin{equation}
    \mathcal{T}(\QTps) \leq 8.49 \:,
\end{equation}
and
\begin{equation}\label{eq:RQTps_bound}
    \mathcal{T}(\RQTps) \leq 7.66 \:.
\end{equation}
Because of \cref{eq:inclusion_QTps_QToi}, it follows that product-state and operational independence lead to the same quantum bound,
\begin{equation}
    \mathcal{T}(\QToi) \leq 8.49 \:.
\end{equation}
On the contrary, the real quantum bounds can differ because of \cref{eq:inclusion_RQTps_RQToi}. It actually follows from \Cref{thm:QToi_subseteq_RQToi} that the maximal value achievable with \RQToi is at least equal to the one achievable with \QToi, 
\begin{equation}
    \mathrm{max}\{\mathcal{T}(\RQToi)\} \geq 8.49 \:,
\end{equation}
since any strategy in \QTps has an equivalent strategy in \RQToi. This highlights that the bound computed by Renou \emph{et al.} relies on the theory-dependent assumption of independent sources associated with \cref{PV_ps}.

\subsection[Real representation of general experiments]{Real representation of general experiments}
The actual main result of Ref.~\cite{hoffreumon2026} is that this inclusion is not only valid for multilocal experiments; rather, it is valid for any experiment that can be modelled within the formalism of quantum circuit theory, provided there is no a priori restriction on the state of the systems before the experiment begins. 

We stress that this theorem only relies on \cref{PI,PII,PIII,PIV} but not on \cref{PV}. As we will see in the next section, it is only in order to falsify its statement that one needs to assume \cref{PV_ps} in conjunction with assuming that the total dimension of the system is bounded.

To explain the theorem we need to recall the quantum instrument formalism: one starts from the usual admission that the most general way to model any conceivable quantum experiment is via a parametrised quantum instrument, which is a family of completely positive and trace-non-increasing (CPTnI) maps $\{\mathcal{M}_{a|x}\}_{a|x}$ indexed by an outcome $a$ conditioned on a setting $x$ and summing up to a CP trace-preserving (CPTP) map $\mathcal{M}_{|x}$. That is, the experiment is modelled by a collection of maps 
\[
    \mathcal{M}_{a|x} \::\: \mathcal{B}(\mathcal{H}^{(in)}) \rightarrow \mathcal{B}(\mathcal{H}^{(out)})
\]
from the initial state of the system prior to the experiment $\rho^{(in)}$ to the final state obtained at the end of the experiment $\rho^{(out)}$. (Here, $\rho^{(in)} \in \mathcal{B}(\mathcal{H}^{(in)})$ where $\mathcal{B}(\mathcal{H}^{(in)})$ indicates the space of bounded operators on $\mathcal{H}^{(in)}$; the dimensions of the input and output systems need not match or be non-trivial.) 
If the outcome $a$ and the setting $x$ were recorded, we have that $\rho^{(out)} = \mathcal{M}_{a|x}(\rho^{(in)})$ and $p(a|x) = \trX{}{\mathcal{M}_{a|x}(\rho^{(in)})}$.

The independence or locality constraints on the experiments are then modelled via a factorisation of each element of the instrument into elements of smaller instruments with some conditions imposed on them, similar to how a multipartite POVM effect factors into a tensor product of several effects, each defined on a subsystem. (The precise way to do so is established by what is known as the quantum circuit formalism---
but it is once again irrelevant for the discussion at hand. Just bear in mind that these factorisations are also able to represent iterative protocols using both classical and quantum side communication.) 

It follows that any quantum experiment is modelled by a set of quantum instruments together with a network topology/a set of causality constraints (i.e., how the various sub-instruments are connected to form the global instrument that models the full experiment). In such a general setting, an experiment is represented as a family of quantum instruments acting on a set of initial systems in an arbitrary state (picture it as the joint vacuum state of all the subsystems before the experimentalists turn on their devices) and producing a set of final systems in the post-measurement state together with the corresponding outcome probability distribution conditioned on settings. 

\subsubsection{Our second theorem}
Our theorem proves that every such quantum protocol has a real quantum analogue provided the systems are associated with suitably larger Hilbert spaces. 
\begin{theorem}[Adaptation of Theorem 2 from Ref.~\cite{hoffreumon2026}]\label{thm:CPTPsubseteqRCPTP}
    Let
    \[\{\mathcal{M}_{a,b_1,b_2,c|x,y_1,y_2,z}\}_{a,b_1,b_2,c|x,y_1,y_2,z} \] 
    be a parametrised quantum instrument acting on an initial quantum state 
    \[\rho^{(in)} \in \mathcal{B}(\mathcal{H}_A^{(in)} \otimes \mathcal{H}_{B_1}^{(in)} \otimes \mathcal{H}_{B_2}^{(in)}\otimes \mathcal{H}_C^{(in)})\:,\] 
    and outputting a post-measurement state
    \[\mathcal{M}_{a,b_1,b_2,c|x,y_1,y_2,z}(\rho^{(in)}) := \rho^{(out)} \in \mathcal{B}(\mathcal{H}_A^{(out)} \otimes \mathcal{H}_{B_1}^{(out)} \otimes \mathcal{H}_{B_2}^{(out)}\otimes \mathcal{H}_C^{(out)})\:,\] 
    such that the instrument and the initial and final states are all described by \QTps. 
    
    Then, there always exists an injective and real-linear function $\mathtt{R}$ from the set of quantum instruments to a set of real quantum instruments 
     \[
     	\{\mathcal{M}_{a,b_1,b_2,c|x,y_1,y_2,z}\}_{a,b_1,b_2,c|x,y_1,y_2,z}  \overset{\mathtt{R}}{\mapsto}\{\mathtt{R}[\mathcal{M}_{a,b_1,b_2,c|x,y_1,y_2,z}]\}_{a,b_1,b_2,c|x,y_1,y_2,z}
     \]
    and an injective and real-linear function $\mathtt{r}$ from the set of quantum states to a set of real quantum states  
    \[
    	\rho^{(in)}\overset{\mathtt{r}}{\mapsto} \mathtt{r}(\rho^{(in)})\:.
    \]
    such that all probabilities are preserved,
    \begin{equation}
    	\trX{}{\mathcal{M}_{a,b_1,b_2,c|x,y_1,y_2,z}(\rho^{(in)})} = \trX{}{\mathtt{R}[\mathcal{M}_{a,b_1,b_2,c|x,y_1,y_2,z}](\mathtt{r}(\rho^{(in)}))} \:;
    \end{equation}
    such that all sequential compositions are preserved, e.g.,
    \begin{subequations}
    \begin{equation}
    	\mathtt{R}[\mathcal{M}_{\vec{o}_2|\vec{s}_2}^{(2)} \circ \mathcal{M}_{\vec{o}_1|\vec{s}_1}^{(1)}] = \mathtt{R}[\mathcal{M}_{\vec{o}_2|\vec{s}_2}^{(2)}] \circ \mathtt{R}[\mathcal{M}_{\vec{o}_1|\vec{s}_1}^{(1)}] ;
    \end{equation}
    \begin{equation}
    	\mathtt{R}[\mathcal{M}_{a,b_1,b_2,c|x,y_1,y_2,z}](\mathtt{r}(\rho^{(in)})) = \mathtt{r}(\mathcal{M}_{a,b_1,b_2,c|x,y_1,y_2,z}(\rho^{(in)})) \:;
    \end{equation}
    \end{subequations}
    and such that all parallel compositions are preserved, e.g.,
    \begin{equation}
    	\mathtt{R}[\mathcal{M}_{a,b_1,b_2|x,y_1,y_2}^{(AB)} \otimes \mathcal{M}_{c|z}^{(C)}] =\mathtt{R'}[\mathcal{M}_{a,b_1,b_2|x,y_1,y_2}^{(AB)}] \otimes \mathtt{R''}[\mathcal{M}_{c|z}^{(C)}] \:,
    \end{equation}
    except if both the initial state and the final state have a product-state factorisation, in which case either:
    \begin{enumerate}
    	\item the initial state cannot be mapped to a real product-state,
    	
        \item[] or
    	
        \item the final state cannot be mapped to a real product-state,
    \end{enumerate}
    but both states are mapped to operationally independent states regardless. 
%
%
%
\end{theorem}
This theorem is a generalisation of \Cref{thm:QT_subseteq_RQT,thm:QToi_subseteq_RQToi} to the instrument formalism. The failure to simultaneously preserve the product-state form of both the initial and final states explains and generalises the separation of \Cref{thm:Renou}. Moreover, the theorem can be enhanced such that the exception (impossibility to map product states to product states) can be overcome in most cases. The bilocal case constitutes one of the few cases where it cannot be removed.

Also, notice how the identification is a real-linear function. Therefore, it cannot induce new hidden or non-linear correlations susceptible to being interpreted as a form of superdeterminism or even as a local hidden-variable model in the real description. 

The theorem, in technical language, precludes the existence of a monoidal functor between \QTps and \RQTps in general. However, it also provides very mild modifications of \RQTps under which the functor may exist.

A consequence of this theorem is that the separation of Renou \emph{et al.} can always be interpreted as a sub-optimal strategy in a larger Hilbert space, and that, in this larger Hilbert space, there always exists a better strategy that reaches at least the same bound as \QTps. This point is detailed next.

\section{Formal exposition of the certification problem}\label{sec:certification}

We come back to $\mathcal{T}(\RQTps) < \mathrm{max}\{\mathcal{T}(\QTps)\}$. Theorem~\ref{thm:Renou} establishes a genuine mathematical separation: within a bilocal experiment modelled according to \RQTps, the swapped-CHSH$_3$ functional is bounded by $7.66$, whereas the \QTps model can reach $6\sqrt{2}\simeq 8.49$. 
Related experiments have since been conducted to check whether the gap between the maximal values of $\mathcal{T}(\QTps)$ and $\mathcal{T}(\RQTps)$ could be witnessed~\cite{Li_2022,Wu_2022}. All of them reported a statistically significant observed value greater than 7.66, with a minimum of $\mathcal{T}_{exp.} = 7.83$ \cite{Wu_2022}. 

It is undeniable that the observed value could not be obtained from a distribution predicted by \RQTps at face value, i.e. of the form of \cref{eq:bilocal_ps}. But how certain can we be that the actual procedure done in the lab could be associated with a distribution of the form \cref{eq:bilocal_ps} to begin with? In other words:

\begin{displayquote}
    On which grounds can bilocal scenarios be associated with a distribution of the form 
    \begin{equation}\tag{\ref{eq:bilocal_ps}}
    p_{|\otimes}(a,b,c|x,z) = \trX{}{(\rho_{AB_1} \otimes \sigma_{B_2C})(A_{a|x} \otimes B_{b} \otimes C_{c|z})}
    \end{equation}
    in \QT and \RQT?
\end{displayquote}

The main debatable points are \cref{PIV,PV} because, when seen as postulates, these are the ones translating the spacelike a priori from outside of the theory into elements of the theory; they dictate how the real-life experimental event ``two independent and spacelike sources distributed among three independent and spacelike measuring parties'' has led to \cref{eq:bilocal_ps}.

\subsection{Issue with postulating an antecedent}
It seems that the question can be disregarded simply by declaring that any consistent mathematical definition can be used as a postulate. That is, by \textit{assuming} that the real-life sources and measurements are represented by operators of the expected form by definition. However, this causes a methodological issue: the antecedent becomes an untestable premise.

This quickly leads to two problems. Firstly, how can an untestable premise lead to a testable conclusion? While the reported value for the swapped-CHSH$_3$ inequality can always be challenged by conducting new experiments or refining the data, there is no way to challenge that $\mathcal{T}(\RQTps)$ could describe bilocal experiments, let alone that $\mathcal{T}(\RQTps)$ describes this particular experiment. Questioning the premise, for example by devising another experiment to test the soundness of \cref{PIV,PV}, becomes forbidden. 
Because of that, it becomes hard to justify the \textit{experimental} aspect of the conclusion. 

For example, stating ``the quantum nature of light is established provided laser light exists in nature'' is not an experimental fact; on the other hand, stating ``the quantum nature of light is established because lasers exist; light sources producing photons with Poissonian statistics have indeed been observed'' is. 

Secondly, where did the antecedent come from if it is untestable? Did von Neumann read on golden tablets in a hat that ``spacelike independence corresponds to tensor product''? Not really. Criteria based on a mix of theory-dependent assumptions backed by experimental evidence were used to assert that spacelike independence is reasonably represented as such in quantum theory. We expect these criteria to be provably applicable to real quantum theory before making the same postulate.

\subsection{Motivating source certification}
Consequently, for the experiment to have a sound description in \QTps or \RQTps, each theory must provide a way to test the premises in order to promote these mathematical definitions into postulates. To do so, the parties can use either theory-dependent or independent observations. 

Renou \emph{et al.} have assumed that the methods to establish \QTps can be transposed to establish \RQTps, modulo the change of field. This assumption is actually correct for the first four postulates. 
For example, by checking that the order of the operations does not influence the observed statistics, or by checking that the statistics are no-signalling, \cref{PIV} can be shown sound regardless of $\kk$. These tests do not prove that the POVMs were exactly the expected ones, but they at least show that any other modification of the postulates than passing from $\kk=\cc$ to $\kk=\rr$ (e.g. allowing entangled or superdeterministic POVM effects to represent a pair of local measurements in \RQT) is either rejected or restricted to have the same predictive power as the theory with the non-modified postulates\footnote{We remark in passing that a failure of \cref{PIV} for $\kk=\rr$ would actually help \RQT to predict larger Bell-inequality violations than \QT. This condition is thus at least as necessary as \cref{PV} to obtain the separation.}.

The ways to test the soundness of \cref{PV_ps} in \QTps, on the other hand, cannot be transferred over the reals. 
The issue originates from a lemma in Ref.~\cite{hoffreumon2026} (Lemma A.14): if the measuring parties are local, they have no means within \RQTps to distinguish preparations satisfying \cref{PV_ps} from \cref{PV_oi}. In particular, local preparations can never be used to prove that the joint state of the four subsystems at the beginning of the experiment was a product state and not an operationally independent but non-product one. Yet, if such a state had been prepared accidentally rather than as a product state, then the experiment would be modelled by a distribution of the form \cref{eq:bilocal_oi}. This is a perfectly valid situation in \RQTps since operationally independent states are, in particular, states, and this kind of experiment does reach the bound of at least 8.49. Hence, the experimental data would have a simple explanation: faulty preparation led the experimentalists to conduct a different experiment than the one proposed by Renou~\emph{et al.}, one in which the sources were not independent in the sense of \cref{PV_ps}. What the thought experiment is doing by observing a violation of the $\mathcal{T} \leq 7.66$ bound, actually, is simply witnessing this faulty preparation (if a RQT explanation is requested).

This problem would not arise in \QTps since, on the one hand, preparations associated with \cref{PV_ps,PV_oi} are equivalent for $\kk=\cc$, and, on the other hand, local tomography is sufficient to conclude that the state at the beginning of the experiment was a product state. 
Consequently, the parties must find a way other than local tomography to reject the faulty preparation explanation.

\subsection{Impossibility of a theory-independent certification of the sources}
The issue is not the validity of \cref{PV_ps} as a mathematical stipulation, but whether experimental evidence establishes that the preparation satisfies its premise. In QT, observing distributions that satisfy operational independence for every experiment involving independent systems can serve as such evidence (provided \ref{PI}--\ref{PIV} have been accepted as postulates). However, in \RQT, operational independence alone cannot establish this:  non-product states can yield the same source-local statistics as product states. These alternative descriptions remain compatible with the observed source-local independence, yet admit the larger set of network correlations described in \Cref{thm:QToi_subseteq_RQToi}. Thus, observing operational independence does not certify the independent preparation to which \cref{PV_ps} assigns a product state.

We could not find another less theory-dependent criterion able to establish \cref{PV} as a postulate in \RQT, or at least one leading to the same predictions as \RQTps. For instance, commutation of the preparation procedures leads to the same increase in the swapped-CHSH$_3$ bound as operational independence because of \Cref{thm:CPTPsubseteqRCPTP}. 
While the question remains open, we can instead consider a theory-dependent certification. In other words, one would prove the starting hypotheses using the other postulates of the theory. In the case at hand, it means establishing, within \RQT, a procedure that certifies that the initial states were prepared independently in the sense of \cref{PV_ps}.

\subsection{Impossibility of theory-dependent certification}

One can imagine that the Renou \emph{et al.} protocol is extended by a pre-experimental stage in which a referee ensures that the sources distributed between the parties are in a product state. In particular, we could allow said referee access to any resource within \RQTps. For example, they may use arbitrarily large entangled ancillae, iterative measure-and-reprepare protocols, joint CPTP maps, etc. 

\Cref{thm:CPTPsubseteqRCPTP} shows that the product-state assumption still cannot be certified without either trusting the dimension of all systems or knowing that the state was a product state to begin with. 
Suppose the Renou \emph{et al.} protocol is extended to include a certification stage between source preparation and measurements. The goal of this stage is to certify that the initial state $\omega^{(in)}$ is indeed of the form $\rho_{AB_1} \otimes \sigma_{B_2C}$. 
Since this stage is associated with an independent referee with unlimited power, it is represented by the most general evolution allowed by \RQTps (here extended to include CPTP maps): a real quantum instrument acting on all subsystems simultaneously. 
Let there be a quantum instrument $\{\mathcal{M}_r\}_r$ associated with two outcomes such that $r=0$ corresponds to the situations where the preparation has succeeded, i.e.,
\begin{equation}
	\mathcal{M}_{r=0}(\omega^{(in)}) = \rho_{AB_1} \otimes \sigma_{B_2C} \:,
\end{equation}
and $r=1$ to those where it has not. This way, the extended protocol is given by 
\begin{equation}
	p(a,b,c,r|x,z) = \trX{}{ ({A}_{a|x} \otimes {B}_{b}\ \otimes {C}_{c|z}) \mathcal{M}_r (\omega^{(in)})} \:,
\end{equation}
and the success of the certification is reduced to the Renou \emph{et al.} protocol,
\begin{equation}\label{eq:rzero}
	p(a,b,c|x,z,r=0) = \trX{}{ ({A}_{a|x} \otimes {B}_{b}\ \otimes {C}_{c|z}) (\rho_{AB_1} \otimes \sigma_{B_2C})} \:,
\end{equation}
with the classical outcome $r=0$ serving as a witness that the preparation succeeded. 

However, one could apply the mappings of \Cref{thm:CPTPsubseteqRCPTP} to \cref{eq:rzero}. In such a case, the experiment now becomes described by a distribution of the form
\begin{equation}
	p(a,b,c|x,z,r=0) = \trX{}{ ({\Tilde{A}}_{a|x} \otimes {\Tilde{B}}_{b}\ \otimes {\Tilde{C}}_{c|z}) (\Tilde{\omega}_{AB_1B_2C})} \:,
\end{equation}
where all operators are still positive real matrices but defined in a larger dimension, and where the certified source is actually an operationally independent state $\Tilde{\omega}_{AB_1B_2C}$. 
In this description, the bound of $\mathcal{T}(\RQTps)\leq 7.66$ has another explanation: the POVMs $\{\Tilde{A}_{a|x}\}_{a|x}$, $\{\Tilde{B}_{b}\}_b$, $\{\Tilde{C}_{c|z}\}_{c|z}$ were not the optimal strategy. The value $6\sqrt{2}\simeq8.49$ can be obtained simply by changing the strategy such that the measurements (and states) are the optimal complex strategy through the preimage of the mappings given by \Cref{thm:CPTPsubseteqRCPTP}. 

In a sense, this theorem guarantees that there are always unseen correlated states in \RQTps that the parties can leverage to obtain the best complex strategy. 
Even when the parties can use any protocol aimed at determining that these states were not present, like with the $\{\mathcal{M}_{r}\}_{r}$ certification stage, the argument can be repeated, i.e., the dimension can always be increased to add other such states. Therefore, for the protocol to be conclusive, the dimensions have to be bounded. However, this cannot happen: the parties have no way of bounding the maximal dimension that their systems can have~\cite{Brunner_2008}. Hence, even tomography of the whole universe cannot escape \Cref{thm:CPTPsubseteqRCPTP}, as it would require knowing its dimension beforehand. 

An interesting observation is that, if the swapped-CHSH$_3$ experiment could itself be described within \RQTps, then the Renou \emph{et al.} thought experiment would provide precisely such an `unspoofable' black-box operation $\{\mathcal{M}_{r}\}_{r}$, which outputs $r=1$ whenever the state is not a product state. Indeed, observing $\mathcal{T}>7.66$ establishes exactly that, with a level of confidence in the black-box operation analogous to that obtained from self-testing the source. The thought experiment could therefore be used to certify its own premise: one could run it during the preparation of the state and use its output to certify that the prepared state is a product state. This procedure could then be iterated, using the same thought experiment to certify the black-box operation used in the previous step, and so on. This `bootstrapping' behaviour illustrates a form of internal consistency of \RQTps: within the mathematical framework, the certification procedure can be recursively grounded in itself. Our results show, however, that this internal bootstrap cannot be established from the experimental data alone: the existence of the required black-box operation cannot itself be certified experimentally.

Summarising, the `independent physical preparations = product-states' assumption associated with \cref{PV_ps} cannot be certified in \RQT. It is thus impossible to use \RQTps to assert that \cref{eq:bilocal_ps} was the proper description of the experiment to begin with. Therefore, if one believes that \RQTps is the proper description of nature, they can always explain the observed data by saying that the physical implementation of the Renou \emph{et al.} protocol has a different description than \cref{eq:bilocal_ps}. 

\end{document}